# Tunable coupling between valley and magnetic excitons in $WS_2$/CrSBr heterostructures

Jianchen Dang[1], Liza Godziashvili[1], Kaichen Xie[2], Arun Ramanathan[3], Vadym Kulichenko[4], Alejandro Busto[1], Jia-Yi Ji[1], Robert Boyd[5], Kenji Watanabe[6], Takashi Taniguchi[7], Sina Najmaei[8], Daniel Rhodes[5], Luis Balicas[9], Xavier Roy[5], Ting Cao[2], Xiao-Xiao Zhang[1*]

[1]*Department of Physics, University of Florida, Gainesville, FL, USA*

[2]*Department of Materials Science and Engineering, University of Washington, Seattle, WA, USA*

[3]*Department of Chemistry, Columbia University, New York, NY, USA*

[4]*National High Magnetic Field Laboratory, Tallahassee, FL, USA*

[5]*Department of Physics, University of Wisconsin Madison, Madison, WI, USA*

[4]*Research Center for Functional Materials, National Institute for Materials Science, 1-1 Namiki, Tsukuba, Japan*

[5]*International Center for Materials Nanoarchitectonics, National Institute for Materials Science, 1-1 Namiki, Tsukuba, Japan*

[8]*U.S. Army DEVCOM Army Research Laboratory, Adelphi, MD, USA*

[9]*Department of Physics and Astronomy, Baylor University, Waco, TX, USA*

**Corresponding Author:* xxzhang@ufl.edu

## Abstract

Interfacial coupling in van der Waals two-dimensional (2D) heterostructures offers an effective route to tailor the electronic, optical, and magnetic properties of materials. Versatile control and precise engineering of such interlayer interactions can be further achieved through external parameters, like electric field, magnetic field, optical excitation, as well as layer stacking geometries. Here, we examined the magneto-optical properties of 2D semiconductor heterostructures made of antiferromagnetic CrSBr and a monolayer $WS_2$, both of which host intriguing excitonic properties. We identified a type-II band alignment between these two layers and the magnetic proximity effects from CrSBr, manifested by an enhanced Zeeman valley splitting in the $WS_2$ layer. The interlayer charge transfer shows strong spin selectivity and is governed by the relative alignment between the valley-spin polarization of photoexcited carriers in $WS_2$ and the magnetization in CrSBr, giving rise to spin-sensitive circularly polarized photoluminescence (PL) from $WS_2$. Furthermore, the highly anisotropic electronic dispersion in CrSBr enables stacking-dependent control of interlayer interactions, as inferred from a twist-angle dependence of PL quenching. Our work establishes a highly tunable 2D system that integrates valleytronics and magnetic excitons through multiple coupling channels, opening up new opportunities in all-optical manipulation of spin and magnon states and for the development of flexible spin-optoelectronic devices.

## Main text

Atomically thin van der Waals crystals are versatile building blocks with diverse physical properties, including a variety of semiconducting, superconducting, magnetic, and topological materials[1-3]. When assembled into hetero- or homostructures, new interfacial interactions can emerge, hosting intriguing properties such as exchange coupling, correlated electronic states, and Moiré Physics[4-6]. The exceptional tunability of van der Waals heterostructures further enables programmable design for compact multifunctional devices for a wide array of applications including optoelectronics, spintronics, sensing and information transduction[5,7,8].

CrSBr has recently emerged as a promising 2D antiferromagnetic semiconductor with anisotropic in-plane magnetic ordering[9,10]. This material attracts wide attention due to its novel magneto-optical effects, arising from a unique coupling between its exciton and magnetic order[11-14]. The exciton wavefunction directly depends on the layer magnetic ordering, which results in large exciton energy shifts with magnetic transitions. Moreover, pulsed optical excitations have also been shown to be an effective method to drive and manipulate long-lived coherent magnon oscillations, revealing multiple magnon modes that couple with excitons[15]. These properties make CrSBr a particularly appealing candidate for integration with 2D monolayer transition metal dichalcogenides[3] (TMDs), which possess rich optical properties and provide full optical control of the valley and spin degrees of freedom. Prior work has demonstrated TMD heterostructures with other 2D magnets, like $CrI_3$ and $Fe_3GeTe_2$, showing strong proximity effects, spin-sensitive light emission, charge transfer, and tunneling[16-20]. However, reports so far in CrSBr heterostructures with $MoSe_2$ and $WSe_2$ showed only limited and unexpected asymmetric coupling behaviors[21-24], which may be attributed to their type-III band alignment. The lack of efficient interlayer interactions with other 2D semiconductors platforms limits their future implementation in versatile spintronics and optoelectronic devices.

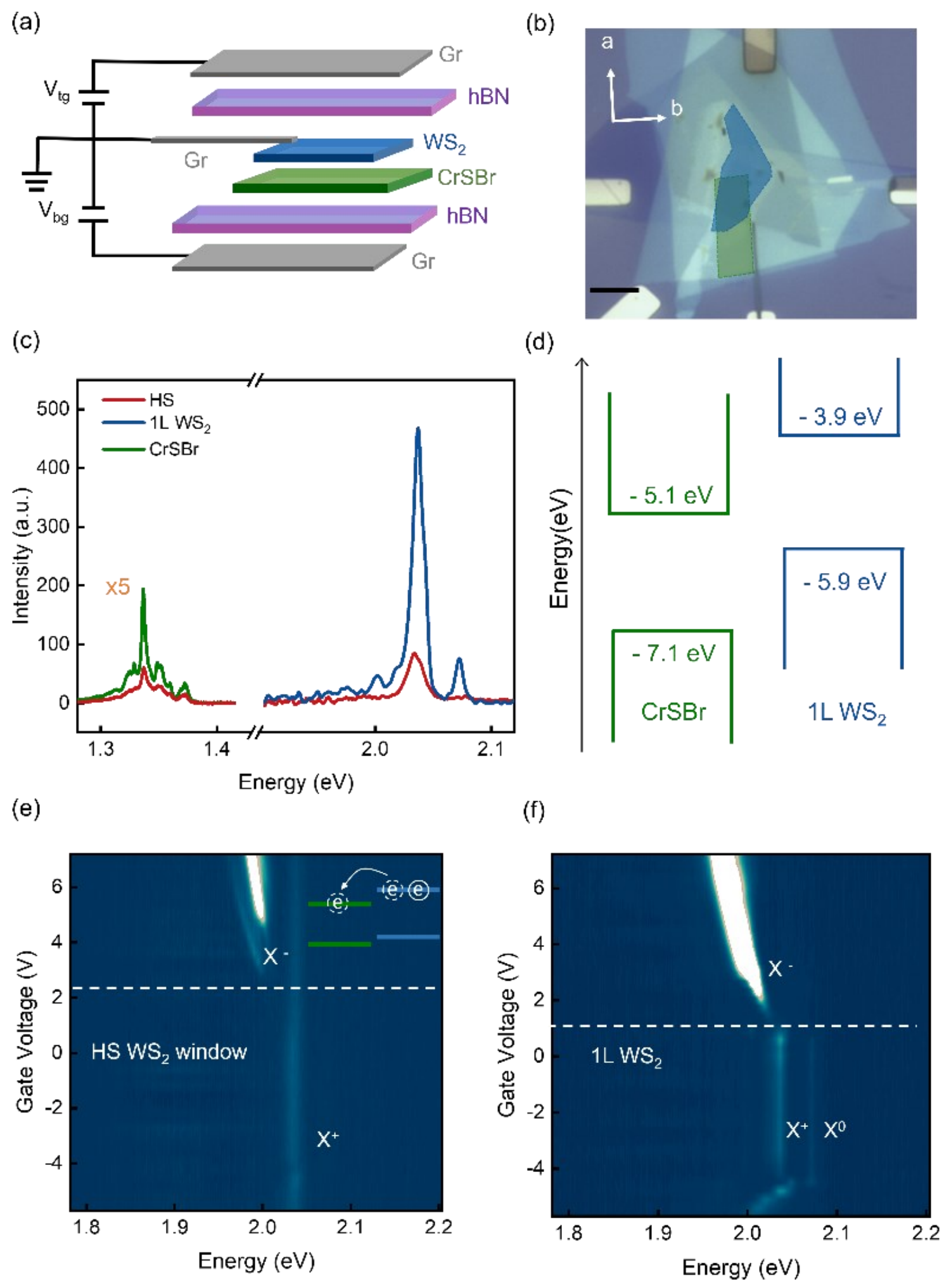

**Figure 1. Structure and optical characteristics of dual-gated $WS_2$/CrSBr heterostructures. a.** Schematic of the dual-gated device architecture. A monolayer $WS_2$ is placed on few-layer CrSBr and electrically grounded via few-layer graphene; thin hBN flakes serve as the dielectric layers, with few-layer graphene also functioning as gate and contact electrodes. **b.** Optical microscope image of a representative device, with the crystallographic a and b axis of the CrSBr flake indicated. The scale bar is 10 µm. **c.** PL spectra from the heterostructure (HS), monolayer $WS_2$, and few-layer CrSBr regions at 4K. The CrSBr window PL signals between 1.3 to 1.4 eV are multiplied by five times for the ease of view with the $WS_2$ window PL. **d.** Energy band diagram representation of the DFT-calculated band alignment between $WS_2$ and CrSBr, revealing a type-II band alignment in a heterostructure. **e.** 4K gate-dependent PL spectra from the $WS_2$ spectra window in the HS region. **f.** Gate-dependent PL spectra from the monolayer $WS_2$ region. $X^0$ is the neutral exciton, $X^+$ and $X^-$ are p-type and n-type trions. The dashed lines in **e** and **f** indicate the onset of n-doping.

In this work, we study the interfacial interactions in monolayer $WS_2$/CrSBr heterostructures and demonstrate a strong valley-spin coupling through both charge transfer and magnetic proximity effects using optical spectroscopies. We identify efficient charge transfer and a type-II band alignment in this heterostructure, consistent with DFT predictions. By analyzing the valley Zeeman splitting in the $WS_2$ layer as a function of an out-of-plane magnetic field, we extract magnetic proximity effects due to the CrSBr magnetic moments with an estimated effective field of ~2.4 T. The circularly polarized photoluminescence (PL) measurements in $WS_2$ show that the population difference in the valley/spin-polarized excitons follows the CrSBr magnetization and thus reveal the spin-selectivity of the interlayer charge transfer. Furthermore, the strong anisotropic crystal structure and quasi-1D nature of the electronic structure in CrSBr provide additional tunability and control of the interlayer interactions. Despite the different lattice geometries between CrSBr and WS2, we observe an interlayer twist angle dependence in the $WS_2$ PL quenching coefficients. The twist dependence of interlayer interactions is inferred to arise from different momentum-matching conditions in these anisotropic heterostructures, providing flexible structural control in heterostructure functionalities. Our results demonstrate highly efficient and tunable interlayer interactions that couple the valley and spin degrees of freedom in CrSBr and $WS_2$ and establish a promising new platform for all-optical and optoelectronic manipulation and readout of spin states for future development of compact spintronics and photonic devices.

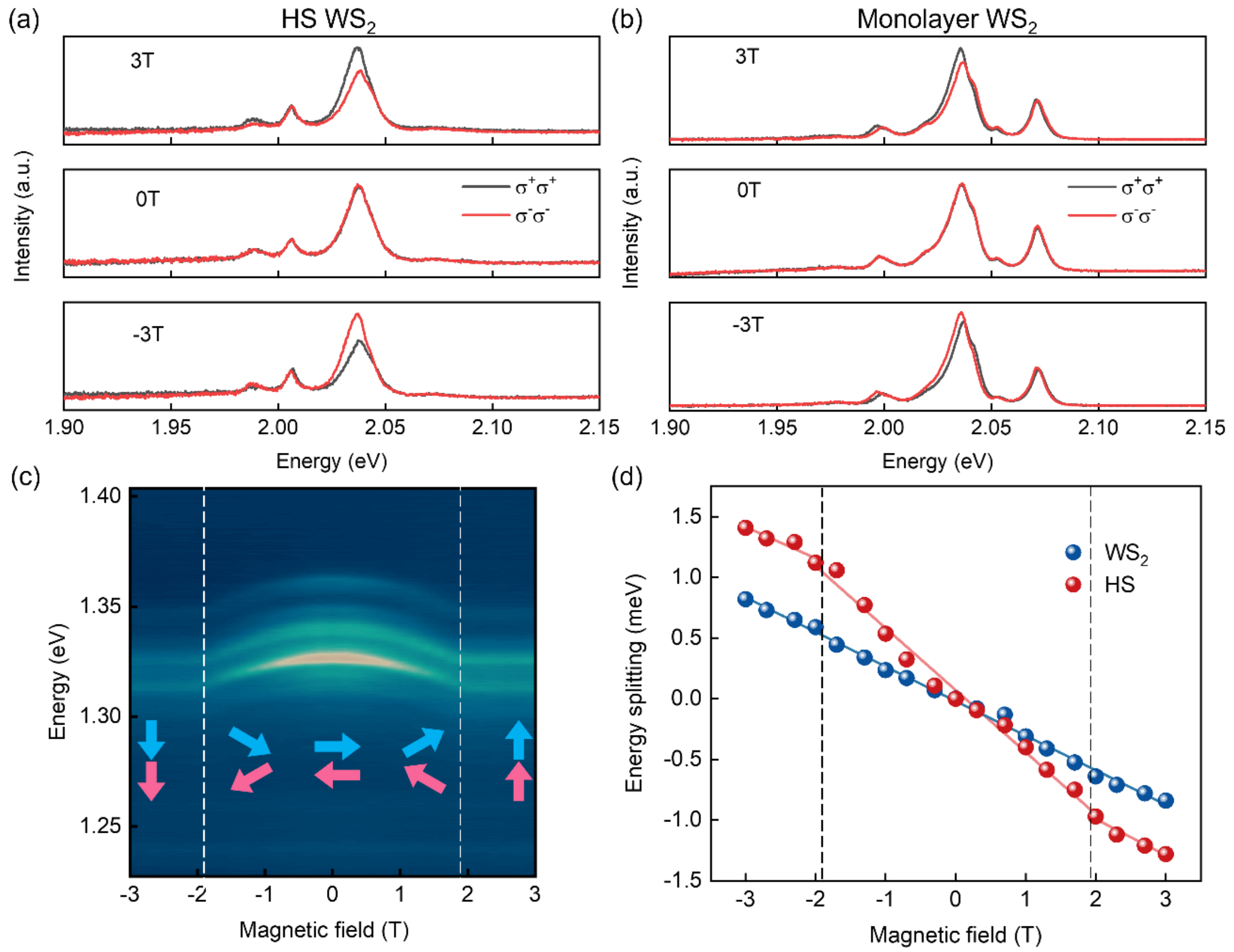

**Figure 2. Magnetic field dependent PL in $WS_2$/CrSBr heterostructures. a,** Circular polarization-resolved PL spectra from the HS region under out-of-plane magnetic fields at ±3T and 0T, in a co-polarized excitation/detection configuration. **b,** Corresponding measurements in the monolayer $WS_2$ region. **c,** Magneto-PL map of CrSBr under out-of-plane magnetic field in HS region, showing a continuous redshift of PL peaks up to a saturation field of ~2T, consistent with a transition from a spin-canted antiferromagnetic state to a ferromagnetic state, as depicted by the blue and pink arrows. **d,** Magnetic field dependence of the trion energy splitting in the HS and 1L WS2. The dashed vertical lines indicate the polarizing field for CrSBr. The straight lines are fitted Zeeman splitting as described in the main text.

Figure 1a plots a schematic of the studied device structure. A monolayer $WS_2$ and a few-layer CrSBr are stacked together through a dry-transfer method and integrated into a dual-gated geometry with hBN and few-layer graphites as the gate dielectrics and electrodes, respectively. Figure 1b shows a white light microscope image of a representative device. The device design incorporates three spatially distinct lateral regions: (i) a monolayer $WS_2$ region, (ii) a $WS_2$/CrSBr heterostructure (HS) area, and (iii) a CrSBr region. This facilitates measurements and comparison across different material configurations. Detailed sample preparation and device fabrication are provided in the Methods section.

Over the HS region, we observe significant PL intensity quenching for both the $WS_2$ and CrSBr PL windows across different devices. Figure 1c compares the PL spectra of the non-heterostructure regions to that of the HS, which shows quenching across different exciton species. Such PL quenching can be considered a hallmark for the formation of a type-II band alignment, as the interlayer transferred charges lead to significant non-radiative loss of exciton populations in both materials. The quenching coefficient in CrSBr is similar for laser excitations above or below the $WS_2$ bandgap. The first-principle density function theory (DFT) calculations of the ionization energy and electron affinity, as shown in Fig. 1d, also confirm our experimental observation of the type-II band alignment between $WS_2$ and CrSBr in different magnetization states. With the DFT-calculated band alignment, we expect a doping shift in the $WS_2$ layer towards heavier p-doping at the HS. This is indeed confirmed in our gate-dependent $WS_2$ PL measurements, which allow us to assign the differences in doping density through intensity differences in neutral excitons and trions[25,26]. As shown in Fig. 1e, the applied gate voltage can tune the monolayer into heavy n-doping at around 1V, while in the HS region, such a transition only occurs around 2.5V, both indicated by the white dashed lines. The doping shift due to the type-II band alignment is estimated to be around $1 \times 10^{12}/cm^2$ with our device geometry using a parallel plate capacitance model. Our results therefore indicate that after optical excitations in both materials, optically-excited electrons in $WS_2$ quickly relax into the CrSBr layer, while the optically-excited holes in CrSBr transfer into $WS_2$.

We further examine how the magnetic moments in CrSBr couple to the valley and spin degrees of freedom in the $WS_2$ layer by estimating the magentic proximity effects. The Zeeman energy splitting between the K and K' valley excitons in TMD materials reflects the local effective magnetic field[3], which will allow us to extract the surface proximity field. To this end, we examined the magnetic field dependence of valley-polarized excitons in $WS_2$ by comparing the co-polarized components with opposite circularly-polarized excitations $\sigma_{\pm}$, which reflects the exciton population and energy in K and K' valleys. The magnetization transitions in CrSBr were monitored from its magneto-PL, as plotted in Fig. 2c. With increasing out-of-plane magnetic field, there is a continuous redshift in CrSBr exciton energy until ~1.8T, which is due to the different exciton hybridization with different layer magnetization. It indicates that in-plane spins cant towards the out-of-plane direction until becoming fully polarized along the out-of-plane direction with the out-of-plane field. The $WS_2$ PL spectra over the HS region and the monolayer region are plotted in Figure 2a and 2 b, which compare the differences when the CrSBr is fully polarized along opposite z-axes. The trion valley splitting in the HS region is significantly larger than that in the monolayer $WS_2$ region. We extracted the trion energy splitting as a function of magnetic field, as plotted in Fig. 2d. The monolayer WS2 region follows a linear field dependence due to its intrinsic valley and spin Zeeman effects. The corresponding slope gives a trion g-factor of ~ 4.8, consistent with previous reports of bright excitons' Zeeman splitting. On the other hand, in the HS region, the K-K' energy splitting shows a larger slope with increasing magnetic field below the saturation fields, with an effective g-factor of ~ 9. Above the saturation fields, the slope falls back to the intrinsic monolayer $WS_2$ g-factor. This enhanced g-factor below the CrSBr saturation arises from the proximity field from the out-of-plane components of the canted spin. It follows $M_Z = M \cdot \sin\theta = M \cdot \frac{H}{H_{pol}}$ and linearly increases with out-of-plane field, where $M$ is the total magnetization after saturation, $\theta$ is the canting angle, and $H_{pol}$ is the polarizing field. Based on the HS region Zeeman splitting, we estimate an effective proximity magentic field of ~ 2.4 T from the fully-polarized CrSBr magnetic moment. We note that the magnetic proximity effects between CrSBr and other TMDs like $MoSe_2$ and $WSe_2$ remain elusive[21-24], with experimental results deviating from the expected in-plane spin canting effects. The mechanism for such contrasting effects in the proximity exchange couplings within the TMD family willrequire future studies.

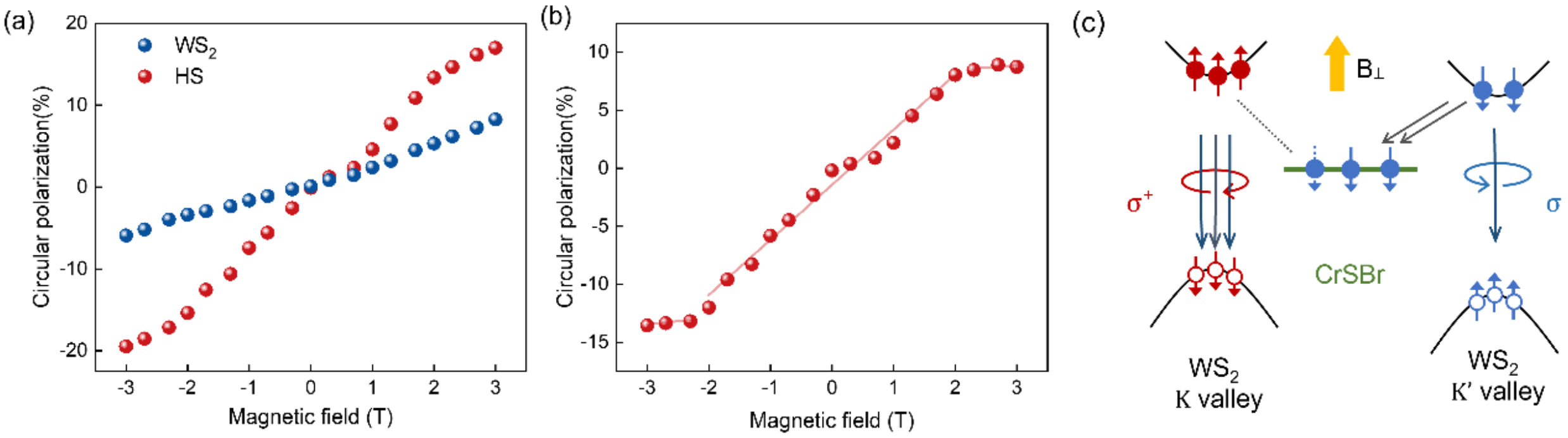

**Figure 3. Enhanced circular polarization in $WS_2$/CrSBr from spin-dependent charge transfer. a,** Extracted circular polarization for the heterostructure (HS) and monolayer $WS_2$ regions against magnetic fields. **b,** Relative polarization obtained by subtracting the monolayer $WS_2$ polarization from that of the HS, showing a field dependence that closely follows the CrSBr magnetic state (Fig. 2c). **c,** Schematic of the Zeeman splitting in $WS_2$ and the spin-dependent charge transfer mechanism in HS. Following this sign convention, CrSBr conduction band spin aligns with the K' valley conduction band spin, and therefore gives rise to higher K valley exciton emission. Only conduction bands involved in optically-allowed transition are plotted.

Apart from exchange-induced magnetic proximity, the interlayer charge transfer process is also shown to be spin-selective. This can be extracted from the valley polarization of $WS_2$ excitons, which reflects the carriers' spin polarization due to the valley-spin locking[27]. As can be directly seen from the valley-polarized trion amplitude differences in Fig. 2a and b at $\pm 3$ T, the HS region develops a larger valley polarization compared to the monolayer $WS_2$ part. Figure 3a plots the comparison of the fitted trion valley polarization as a function of the magnetic field. The circular polarization here is defined as $(I_{\sigma^+\sigma^+} - I_{\sigma^-\sigma^-})/(I_{\sigma^+\sigma^+} + I_{\sigma^-\sigma^-})$. The linear increase in valley polarization in monolayer $WS_2$ arises from the intrinsic Zeeman effects with an out-of-plane magnetic field, which creates valley population difference via the Zeeman splitting (Fig. 3a). To isolate the contribution from the interlayer charge transfer, we subtract the monolayer $WS_2$ valley polarization from the HS region, which is plotted in Fig. 3b. The polarization shows an expected linear field dependence as the spins in CrSBr are canting towards the out-of-plane direction with the external field and saturates when the spins are fully-polarized. This can be understood by considering their type-II band alignment and the corresponding spin-dependent charge transfer rate. When the out-of-plane spin components in CrSBr (after being polarized along the out-of-plane direction) aligns/anti-align with the $WS_2$ K/K' valley spin indices, there is more efficient/weaker interlayer transfer for the valley-polarized electrons in $WS_2$ (Fig. 3c). The magnetically induced valley imbalance leads to the observed valley polarization in the PL. We note that over the HS region, the magnetically quenched trions correspond to the valley with a higher emission energy under the Zeeman splitting (Fig. 2a). This is also consistent with our expectation. When the magnetic field is pointing along the + *z* direction, the K and K' valley conduction and valence bands shifts according to their combined spin, valley and orbital magnetic moments, which will induce an energy blue shift in the K' valley[28] following the spin orientation convention shown in Fig. 3c. Correspondingly, under this +*z* external magnetic field (above the polarizing field), the CrSBr conduction band electrons will be polarized along the -*z* direction, which will assist the interlayer conduction band electron transfer from the K' valley and give rise to larger quenching in the K' valley, which has a higher exciton energy. Notably, the optically excited spin/valley-polarized carriers in $WS_2$ are expected to relax to the lowest conduction bands of CrSBr, which possess highly anisotropic dispersions along the crystalline a- and b-axes. Despite the differences in momentum relaxation during the interlayer charge transfer, which may contribute to spin depolarization, we observed consistent spin-dependent quenching for heterostructures with different crystalline alignments, establishing it as a robust platform for interlayer spin and valley coupling and manipulation.

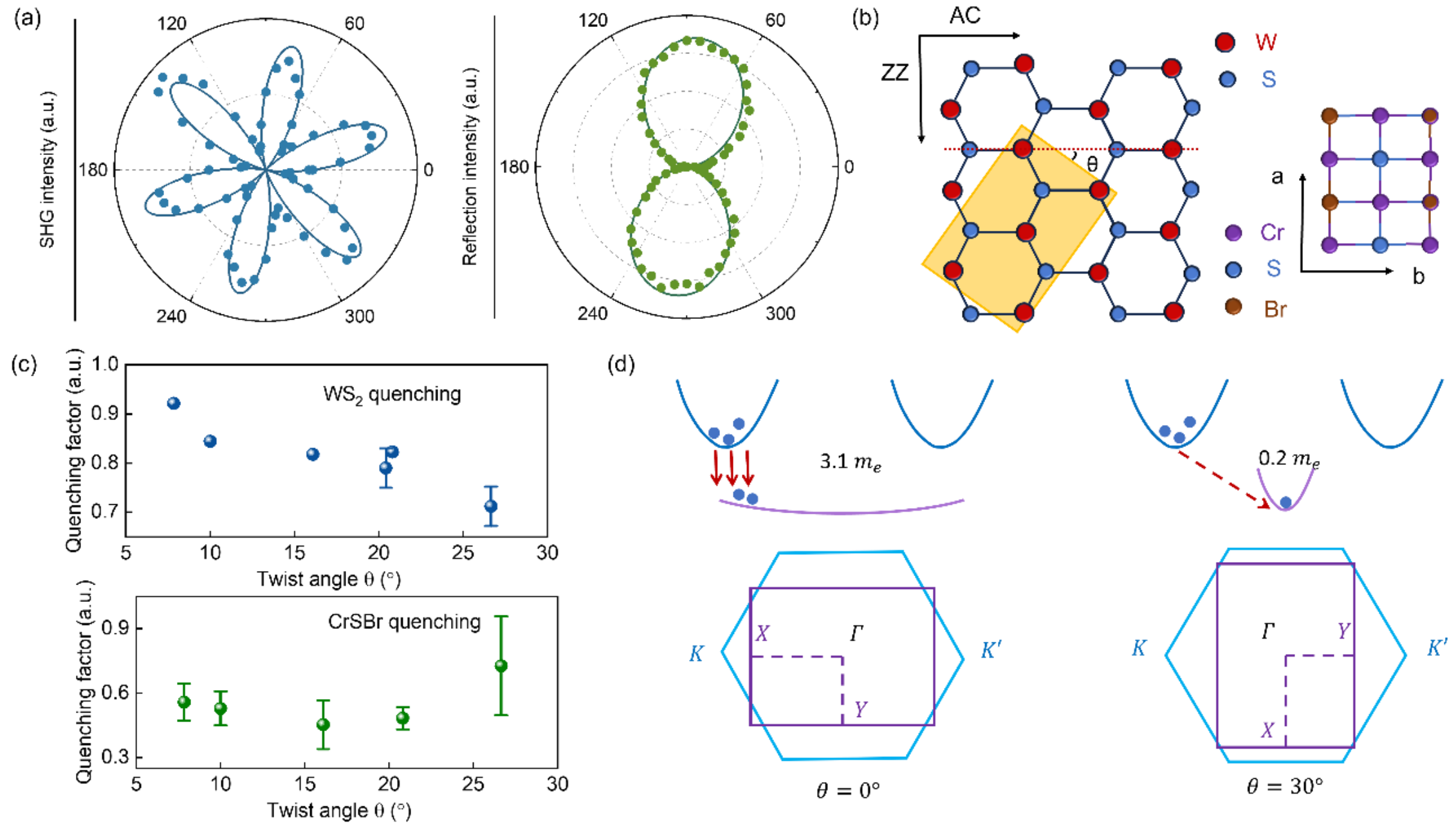

**Figure 4. Twist-angle dependent interlayer interactions. a,** Polarization-resolved second-harmonic generation (SHG) pattern from monolayer $WS_2$, showing a six-petal symmetry characteristic of the hexagonal lattice (left). On the right is the polarization-resolved reflectivity of CrSBr, where the reflectivity maximum aligns with the crystallographic a-axis due to the strong in-plane anisotropy. **b,** Schematic of the in-plane lattice arrangement (top view), illustrating the hexagonal $WS_2$ lattice stacked on the orthorhombic CrSBr lattice. Twist angle θ is defined between the $WS_2$ armchair direction and the CrSBr b-axis. **c,** Twist-angle dependence of the quenching coefficients in the $WS_2$ window (upper panel) and in the CrSBr window (lower panel). The error bars indicate the spatial inhomogeneity measured within the different sample locations. **d,** Schematic heterostructure band alignment when $\theta = 0°$ (left) and 30° (right). The bottom panels plots the corresponding first Brillouin zones of CrSBr (purple) and $WS_2$ (blue), while the top panels depict their conduction bands along $K - \Gamma - K'$ direction. When $\theta = 0°$, the transfer of interlayer conduction band electrons into the flat CrSBr band becomes efficient. When $\theta = 30°$, additional phonon emissions are needed to satisfy the momentum conservation.

The highly anisotropic bands in CrSBr offer additional twist tunability[29,30] for the interlayer coupling. The rectangular CrSBr and honeycomb $WS_2$ lattices have distinctly different geometries, and therefore a commensurate moiré potential modulation is not expected. However, the anisotropic CrSBr lattice breaks the three-fold rotation in monolayer $WS_2$ and offers structural control of the heterostructure functionalities[31]. Upon photoexcitation in $WS_2$, excited carriers populate its $K$ and $K'$ valleys, which will relax into CrSBr due to the type-II band alignment while emitting phonons to satisfy the momentum conservation. Momentum matching conditions at different stacking alignments[32] will therefore affect the charge transfer process. To examine this possibility, we fabricated and measured CrSBr/$WS_2$ devices with different twisting angles and compared their PL quenching efficiency to monitor the interlayer modulation. Their crystalline axis were determined by second harmonic generation[33] ($WS_2$) and linear reflectivity[11] (CrSBr), as shown in Fig. 4a. Here we define the twist angle $\theta$ as the difference in angles between the armchair direction of $WS_2$ and the b-axis of CrSBr (Fig. 4b). We compared devices with twisting angles ranging from $\theta = 0°$ to 30°, corresponding to when the b-axis aligns with the $WS_2$ armchair and zigzag directions, respectively. As shown in Fig. 4c, the twist angle-dependent $WS_2$ PL quenching shows an overall increase when $\theta$ decreases to 0°. In comparison, PL quenching in CrSBr shows no obvious twist angle dependence within the range of error bars. We note that these PL quenching efficiencies are sensitive to extrinsic device fabrication processes, but most of our examined devices follow this trend. Additional PL and reflectivity characterizations of other devices are provided in supplementary figures. To understand this trend, we compare the first Brillouin zone overlap between CrSBr and $WS_2$. As sketched in Fig.4d, when the twist angle $\theta = 0$, $K - K'$ in $WS_2$ aligns with the $\Gamma - K$ in CrSBr. In this configuration, CrSBr possesses a mostly flat (large effective mass) conduction band, which imposes little momentum constraint during the interlayer charge transfer and can lead to strong quenching. On the other hand, when $\theta = 30°$, the $K - K'$ direction now aligns with $\Gamma - Y$ with a much smaller conduction band effective mass. At this twist angle, the conduction band minima of the heterostructure lie in the CrSBr $\Gamma$ point. The photoexcited carriers in $WS_2$ will need to overcome the large momentum mismatch from $K$ to $\Gamma$, which can lead to the observed slower charge transfer rate and lower quenching ratio. In contrast, when we consider the PL quenching in CrSBr, the exciton population decay arises from the CrSBr valence band holes transferring into the $WS_2$ layer. The Γ point in a monolayer $WS_2$ is also a local valence band energy maxima and only slightly lower than the $K/K'$ valleys. We therefore expect efficient $\Gamma - \Gamma$ charge transfer from CrSBr to $WS_2$ regardless of the twist angle, which is consistent with our observations.

**Discussion**

We demonstrated efficient interlayer modulations between van der Waals semiconductor heterostructures of magnetic CrSBr and monolayer $WS_2$, including the exchange-induced proximity effects and interlayer charge transfer with a type-II band alignment. Our results show that spin polarization is preserved during the interlayer charge transfer process, providing van der Waals interface access to coherent magnons and magnon-coupled excitons in CrSBr[13-15] for spin transport and information encoding[8,34,35]. By utilizing the valley-spin coupling and valley selection rules in this heterostructure, we demonstrated the possibilities of magnetization sensing through valley-polarized exciton states, all-optical manipulation of spin and magnon, and future spintronics applications in optoelectronics. In particular, we revealed efficient structural modulation of the interlayer coupling by varying the relative twist angles between the $WS_2$ monolayer and CrSBr, which is a result of the highly anisotropic conduction bands of CrSBr. Such twisting-angle tunability adds deterministic control to manipulate and utilize interlayer interactions in 2D magnetic heterostructures. Our findings open up new opportunities to design anisotropic photonic devices that leverage both the valleytronics of TMDs and the coherent magnon states in CrSBr with optical, electronic, and structural tuning knobs.

# Methods

**Material synthesis**

CrSBr single crystal synthesis from A.R. and X.R. follows the previously established procedure[36]. The following reagents were used as received unless otherwise stated: chromium powder (99.94%, 200 mesh, Alfa Aesar), sulfur pieces (99.9995%, Alfa Aesar), bromine (99.99%, Aldrich) and chromium dichloride (anhydrous, 99.9%, Strem Chemicals). To begin, $CrBr_3$ was synthesized from Cr (1.78 g, 34.2 mmol) and $Br_2$ (8.41 g, 52.6 mmol) with one end of the tube held at 1,000 °C and the other end at 50 °C with a water bath. Note that one end of the tube must be maintained below 120 °C to prevent overpressure. Chromium (0.174 g, 3.35 mmol), sulfur (0.196 g, 6.11 mmol) and $CrBr_3$ (0.803 g, 2.75 mmol) were loaded into a fused silica tube 12.7 mm outer diameter; 10.5 mm inner diameter. The tube was evacuated to ~30 mTorr and flame sealed to a length of 20 cm. It was then was placed into a computer-controlled, two-zone, tube furnace. The source side was heated to 850 °C over 24 h, allowed to soak for 24 h, heated to 950 °C over 12 h, allowed to soak for 48 h and then cooled to ambient temperature over 6 h. The sink side was heated to 950 °C over 24 h, allowed to soak for 24 h, heated to 850 °C over 12 h, allowed to soak for 48 h and then cooled to ambient temperature over 6 h. The crystals were cleaned by soaking in a 1 mg $ml^{-1}$ $CrCl_2$ aqueous solution for 1 h at ambient temperature. After soaking, the solution was decanted and the crystals were thoroughly rinsed with deionized water and acetone. Residual sulfur residue was removed by washing with warm toluene.

Single crystals of CrSBr from L.B. and V.K. were synthesized via chemical vapor transport (CVT). High-purity chromium powder (60 mesh, 99.999%), sulfur pieces (99.999%), and liquid bromine (99.999%) were loaded into a quartz tube (inner diameter 16 mm, length 250 mm) in a 1:1:1.05 molar ratio. To prevent premature reaction during evacuation, the mixture was cooled down to liquid nitrogen temperatures prior to sealing under high vacuum. The reactants were pre-reacted by maintaining one end of the tube at 500 ºC while keeping the opposite end at room temperature for 24 hours. The tube was subsequently transferred to a two-zone horizontal furnace. To clean the growth region, a reverse temperature gradient was initially established: the source zone was heated up to 800 ºC and the growth zone up to 860 ºC for 24 hours. The source zone was then ramped to 940ºC over 6 hours, establishing a thermal gradient of 940 ºC → 860 ºC (source to growth zone). These conditions were maintained for two weeks to foster single-crystal growth. The resulting crystals were harvested and washed sequentially in acetone and hot toluene.

Single crystals of $WS_2$ were done via a salt-flux transport approach[37]. A stoichiometric ratio of W powder (99.999%) and S pieces (99.999%) was loaded into a quartz ampoule (16 mm outer diameter, 12 mm inner diameter), along with 30 mg of excess S. Then a eutectic mixture of NaCl (99.998%) and CsCl (99.999%), approximately 7 g total, was loaded on top of the starting precursors under an argon environment and the ampoule was sealed under vacuum. The ampoule was then heated horizontally in a tube furnace up to 1100 °C (hot zone), 1000 °C (cold zone), over one day and dwelled there for 3 weeks. Afterwards, the sample was cooled to 500 °C at 1.25 °C/h and then the furnace was shut off. $WS_2$ single crystals were removed from the encapsulating salt flux by immersion in water.

**Device fabrication.**
Few-layer graphite, hexagonal boron nitride (hBN), monolayer $WS_2$, and few-layer CrSBr flakes were mechanically exfoliated from bulk crystals and identified by optical contrast under an optical microscope. The exfoliation of CrSBr was performed inside an inert nitrogen-filled glovebox to prevent environmental degradation. The van der Waals heterostructure was assembled using a dry-transfer technique with a polycarbonate (PC) stamp and subsequently released onto a substrate with prepatterned gold electrodes. The transfer steps were conducted in a nitrogen-filled glove box.

**Optical spectroscopy measurements.**
The devices were mounted onto 3D piezoelectric stage in an optical cryostat (attoDry1000) with a base temperature of 4 K. The cryostat was equipped with a superconducting solenoid magnet capable of providing magnetic fields ranging from −9 T to 9 T.  For PL measurements, the emitted signal was collected through an objective lens with a numerical aperture (NA) of 0.82 and analyzed using a grating spectrometer coupled to a CCD detector (Princeton Instruments SpectraPro HRS300 + PIXIS). The polarization of the emission was analyzed using a quarter-wave plate followed by a linear polarizer. Optical excitation was provided by a 540 nm ultrafast pulsed laser with a focal spot diameter of approximately ~1 μm. SHG measurements were performed using a 780 nm femtosecond laser Menlo Systems C-Fiber 780, with a repetition rate of 100 MHz and a pulse duration of < 200 fs. In polarization-resolved reflectivity measurements, a 633 nm continuous-wave He–Ne laser was used. The polarization state of the incident light was controlled by rotating a half-wave plate placed in the excitation path. The reflected light from the sample was collected by the objective lens and detected by the spectrometer/CCD system. By continuously rotating the half-wave plate, the reflectivity of the sample was recorded as a function of the incident polarization angle.

### DFT calculations

Ab initio calculations were performed using DFT implemented in the VASP package. PAW pseudopotentials with a plane-wave energy cutoff of 430 eV were employed. We employed the spin-polarized Perdew–Burke–Ernzerhof exchange–correlation functional for structural relaxation, and HSE06 hybrid function with both spin-orbit coupling and spin polarizations included for calculations of electron affinity and ionization potential of monolayer $WS_2$ and CrSBr. The out-of-plane unit cell lattice parameter is fixed at 20 Å to avoid interactions between periodic images.

## Data availability

The data that support the findings of this study are available within the paper and its Supplementary Information. Additional data are available from the corresponding authors upon request.

## Acknowledgment
X-X.Z. acknowledge the support from the Department of Energy (DOE) award DE-SC0022983. This work was partly conducted at the Research Service Centers of the Herbert Wertheim College of Engineering at the University of Florida. L.B. and V.K. are supported by the US DoE, BES program through award DE-SC0002613. The National High Magnetic Field Laboratory acknowledges support from the US-NSF Cooperative agreement Grant DMR-2128556, and the State of Florida. Synthesis work at Columbia University was supported by the Air Force Office of Scientific Research (AFOSR) under MURI grant number A9550-25-1-0288. D.R. acknowledges support from the Office of Naval Research (N000142512276). R.B. acknowledges support from the DEVCOM Army Research Laboratory (ARL) Research Associateship Program (RAP), administered by Oak Ridge Associated Universities (ORAU) for ARL through Cooperative Agreement (CA) W911NF-24-2-0054. The views and conclusions contained in this document are those of the authors and should not be interpreted as representing the official policies, either expressed or implied, of the Army Research Laboratory or the U.S. Government. The U.S. Government is authorized to reproduce and distribute reprints for Government purposes notwithstanding any copyright notation herein.

## Author contributions
X-X.Z. and J. D. designed the study. J.D. and L.G. fabricated the device, performed the magento-optical measurements and data analysis. A.B. and J.-Y. J. performed twist angle measurements. K.W. and T.T. grew the bulk hBN crystals. CrSBr single crystal growth at Baylor University/NHMFL is carried out by V.K., L.B.. CrSBr crystals from Columbia University is grown A.R., under the supervision of X.R.. D.R. and S.N supervised the single crystal synthesis of $WS_2$ by R.B. X-X.Z., J.D. and L.G. wrote the initial manuscript draft. All authors discussed the results and commented on the manuscript.

## Competing Interests
The authors declare no competing interests.